\documentclass[12pt]{article}
\def\half{{\textstyle{1\over2}}}
\def\2{{1\over 2}}

\newcommand{\rf}[1]{(\ref{#1})}
\def\b{\bar}

\renewcommand{\t}{\tilde}
\newcommand{\p}{\partial}
\newcommand{\bp}{\bar{\partial}}
\newcommand{\tc}{\tilde c }
\usepackage{amssymb}

\title{
\bf{Perturbed Beta-Gamma Systems and Complex Geometry}}
\author{Anton M. Zeitlin\footnote{anton.zeitlin@yale.edu http://pantheon.yale.edu/\~ az84 http://www.ipme.ru/zam.html}   
\footnote{On leave of absence from the St.-Petersburg Division of Steklov Mathematical Institute}\\
Department of Mathematics,\\
Yale University,\\
442 Dunham Lab, 10 Hillhouse Ave\\
New Haven, CT 06511}
\date{}

\begin{document}
\maketitle
\begin{abstract}
We consider the equations, arising as the conformal invariance conditions
of the perturbed curved beta-gamma system. These equations have the physical meaning
of Einstein equations with a B-field and a dilaton on a hermitian manifold, where the 
B-field 2-form is imaginary and 
proportional to the canonical form associated with hermitian metric.
We show that they decompose into linear and bilinear equations and lead to the vanishing 
of the first Chern class of the manifold where the system is defined. 
We discuss the relation of these equations to the generalized Maurer-Cartan structures related to 
BRST operator. Finally we describe the relations of the generalized Maurer-Cartan bilinear operation
 and the Courant/Dorfman brackets.

\end{abstract}

\section{Introduction: Perturbations of $\beta$-$\gamma$ Systems and Sigma Models}
The so-called $\beta$-$\gamma$ (beta-gamma) systems recently drew a lot of attention. 
The interest to them started in the mathematics papers \cite{schekhtman},\cite{malikov} where 
they were considered in the context of the sheaves of vertex operators 
and chiral de Rham complex. Soon after that, the obtained results appeared to be very useful 
in applications to string physics, e.g. topological strings \cite{kap}, the geometry of (2,0)
 models \cite{witten2}, mirror symmetry \cite{frenkel},  
and the pure spinor superstring formalism \cite{berkovits}, \cite{nekrasov}.
 
The $\beta$-$\gamma$ system is a two dimensional conformal field theory with the action:
\begin{eqnarray}
S_0=\frac{1}{2\pi h}\int_\Sigma d^2 z \beta_i\bar{\partial} \gamma^{i},
\end{eqnarray}
where $\beta_i$ and $\gamma^{j}$ are of conformal weights (1,0) and (0,0) correspondingly 
and $h$ is some constant. The action above 
leads to the following operator product between these two fields:
\begin{eqnarray}
\gamma^{i}(z_1)\beta_{j}(z_2)\sim\frac{h\delta^{i}_{j}}{z_1-z_2}.
\end{eqnarray}
This model looks very simple, however, if we take into account the geometric aspects, 
namely treating $\beta_i(z)dz$ as the sections of 
$\gamma^*({T'}^*M)\otimes{T'}^*{\Sigma}$, where $\gamma$ describes a map from Riemann 
surface $\Sigma$ to the complex manifold $M$ and prime denotes the 
holomorphic part of the appropriate cotangent bundle, the $\beta$-$\gamma$ system becomes highly nontrivial.
 For example, one can ask a question: how to keep operator 
products preserved after coordinate transformations or how to keep the appropriate energy momentum tensor,
 well defined on the intersections of the appropriate 
coordinate patches (for a nice review see \cite{nekrasov}). The answer to the first question
 is nontrivial but the second one is quite simple. In order to be unaffected under
 the coordinate change, the energy-momentum tensor should 
be modified from the original one, $T_0=-h^{-1}\beta_i\p \gamma ^i$ to the following:
\begin{eqnarray}\label{t}
T=-h^{-1}\beta_i\p \gamma ^i-1/2\p^2\log\omega(\gamma),
\end{eqnarray}    
where $\omega$ is the density function for the holomorphic top form on the manifold $M$.
 This leads to a globally 
defined conformal invariant theory. An obvious requirement for the possibility of such modification is that 
this top form related to $\omega$ should be nonvanishing, or in other words the manifold $M$
 should have the vanishing first Chern class \cite{witten2}, \cite{nekrasov}. 

The inclusion of the 
additional dilatonic term in the energy momentum tensor \rf{t} yields the modification of the action \cite{fts}:
\begin{eqnarray}
S_0=\frac{1}{2\pi h}\int_\Sigma d^2 z (\beta_i\bar{\partial} \gamma^{i}+\frac{1}{4}h\sqrt{s}R^{(2)}(s)\log\omega(\gamma)),
\end{eqnarray}
where $R^{(2)}$ is the curvature of the worldsheet metric $s_{ab}$. It is worth noting also that this
 term entering the action is the secondary characteristic class 
associated with class $c_1(M)c_1(\Sigma)$. 

It appears that the $\beta$-$\gamma$ system considered above can be treated as an infinite-radius
 limit of the certain nonlinear sigma-model \cite{nekrasov}, \cite{lmz}. Really, let's consider 
the action for the sigma model
\begin{eqnarray}\label{sigma}
\mathcal{S}=\frac{1}{4\pi h}\int_\Sigma d^2 z((G_{\mu\nu}+B_{\mu\nu})\partial X^{\mu}\bar{\partial}X^{\nu}+h\sqrt{s}R^{(2)}(s)\Phi).
\end{eqnarray}
We impose several conditions on the metric and the $B$-field. First of all we require 
the metric to be hermitian in the certain system of 
coordinates: $G_{ik}=G_{\b i\b k}=0$, $G_{i\b k}\equiv g_{i\b k}$. The other conditions concern $B$-field:   
\begin{eqnarray}\label{gbcond}
g_{i\b k}+B_{i\b k}=0,\quad B_{ik}=B_{\b i\b k}=0.
\end{eqnarray}
This means that the action \rf{sigma} can be rewritten as follows:
\begin{eqnarray}\label{sigmafo}
\mathcal{S}=\frac{1}{2\pi h}\int_\Sigma d^2z
(g_{i\bar{j}}\bar{\partial} X^i\partial X^{\bar{j}}+1/2h\sqrt{s}R^{(2)}(s)\Phi)
\end{eqnarray}
and via the first order formalism, this sigma model can be transformed \cite{lmz} to the first order one:
\begin{eqnarray}\label{fo}\label{bgp}
S_{g}= \frac{1}{2\pi h}\int_\Sigma d^2 z (p_i\bar{\partial} X^{i}+
p_{\bar{i}}{\partial} X^{\bar{i}}-g^{i\bar{j}}p_i p_{\bar{j}} +1/2h\sqrt{s}R^{(2)}(s)\Phi_0),
\end{eqnarray}
where $\Phi_0=\Phi-\log\sqrt {g}$. The infinite radius limit of the sigma model therefore corresponds to the action above without
 the perturbation operator $g^{i\bar{j}}p_i p_{\bar{j}}$. Obviously, the identification
 of the resulting model with the conformally invariant, 
well defined $\beta$ -$\gamma$ system and its complex conjugate, can be done by means of 
the identification $\beta_i=p_i$ and $\gamma^i=X^i$ 
but only in the case if $\Phi_0$ decomposes in the sum of holomorphic and antiholomorphic functions. 

In section 2, we examine the geometric and algebraic properties of the conformal invariance conditions of 
the sigma model (i.e. beta-function) of the zero and first order in $h$ of the theory with action \rf{sigmafo}. 
 These conditions are the Einstein equations with a B-field and a dilaton \cite{eeq},\cite{Polbook} equipped with constraints \rf{gbcond}: 
\begin{eqnarray}
&&R_{\mu\nu}=-{1\over 4} H_{\mu\lambda\rho}H_{\nu}^{\lambda\rho}+2\nabla_{\mu}
\nabla_{\nu}\Phi,\nonumber\\
&&\label{e2}
\nabla_{\mu}H^{\mu\nu\rho}-2(\nabla_{\lambda}\Phi)H^{\lambda\nu\rho}=0,\nonumber\\
&&
\label{e3}
4(\nabla_{\mu}\Phi)^2-4\nabla_{\mu}\nabla^{\mu}\Phi+
R+{1\over 12} H_{\mu\nu\rho}H^{\mu\nu\rho}=0.
\end{eqnarray}
Expressing them in terms of $g^{i\b j}$ and $\Phi_0$, we find out that one of the equations is simply
\begin{eqnarray}
\p_i\p_{\b k}\Phi_0=0,
\end{eqnarray}
what precisely means that $\Phi_0$ is given by the sum of holomorphic and antiholomorphic terms 
which by construction lead to the existence of nonvanishing 
holomorphic top form and therefore to the vanishing of the first Chern class of the manifold. 
The equations on $g^{i\b j}$ are either linear or $bilinear$. 

In section 3, we discuss another algebraic structure, rudiments of which were given in \cite{lmz}, \cite{zeit}. Namely, in \cite{lmz}   
there was introduced a conjecture (at first introduced by A.S. Losev \cite{losev}), 
 that the conformal invariance condition for the perturbed $\beta$-$\gamma$ system \rf{bgp} can be expressed as follows:
\begin{eqnarray}\label{gmcint}
C_1(\phi^{(0)})+C_2(\phi^{(0)},\phi^{(0)})+C_3(\phi^{(0)},\phi^{(0)},\phi^{(0)} )+...=0.
\end{eqnarray}
Here $C_n$ are graded symmetric multilinear operations satisfying special quad\-ra\-tic relations generating $L_{\infty}$-like structure \cite{zwiebach},\cite{stasheff}, 
such that 
$C_1$$(\phi^{(0)})$ =\\$[Q_{\omega}$$,$$ \phi^{(0)}]$, where $Q_{\omega}$ 
is the usual BRST operator \cite{brst}, \cite{pol} associated with 
energy-momentum tensor \rf{t}, and $\phi^{(0)}$ 
is a differential polynomial in $c, \tc $ ghost fields, of ghost number 2 and 
$[b_{-1},[\t b_{-1},\phi^{(0)}]]$ = $h^{-1}g^{i\b j}p_ip_{\b j}$. We  explicitly construct 
the $C_2$ operation and expanding 
$\phi^{(0)}=\sum^{\infty}_{n=1}t^n\phi_n^{(0)}$ by the formal parameter $t$, we find the agreement 
between \rf{gmcint} and the conformal invariance conditions 
on $g^{i\b j}, \Phi_{0}$ we have derived before, up to the order $t^2$. Moreover, we expect that the symmetries of \rf{gmcint} have the form:
\begin{eqnarray}
\delta\phi^{(0)}=\varepsilon(C_1(\xi^{(0)})+C_2(\phi^{(0)},\xi^{(0)})+
C_3(\phi^{(0)},\phi^{(0)},\xi^{(0)} )+...),
\end{eqnarray} 
where $\varepsilon$ is infinitesimal and $\xi^{(0)}$ is of ghost number 1. 
In the last part of subsection 3.2., we obtain that the symmetry under holomorphic transformations of the 
equations on $g^{i\b j}, \Phi_0$ can be obtained in such a way. 
In subsection 3.3., we indicate some important properties of the bilinear operation $C_2$ related to 
pure chiral data. Namely, we derive the relation of $C_2$ to Courant \cite{courant} and 
Dorfman \cite{dorfman} brackets. 

In Conclusions we underline some important features of the above formalism and outline the ways of further development.

\section{Geometric Aspects of Conformal Invariance Conditions}

In this section, we  show that the conformal invariance conditions of the perturbed $\beta$-$\gamma$ 
system (i.e. Einstein equations) reduce to the bilinear system of equations on $g^{i\b j}$. First of all, we define some algebraic operations on the sections 
$\Gamma(T'M\otimes T''M)$, where we denoted by $T'M$ and 
$T''M$ holomorphic and antiholomorphic tangent bundle $TM=T'M\oplus T''M$ correspondingly.

Suppose $\mathbf{g}=g^{i\b j}\p_i\otimes\p_{\b j}$, $\mathbf{h}=h^{i\b j}\p_i\otimes\p_{\b j}$ 
are bivector fields on some complex manifold 
$M$. If one expands in some neighborhood $g=\sum_I v^I\otimes\b{v}^I$, $h=\sum_J w^J\otimes\b{w}^J$ 
(sum over $I$ and $J$ can be possibly infinite), where $v^I, w^J$ 
($\b v^I, \b w^J$) are (anti)holomorphic sections of $T'M$ ($T''M$), one can construct a new bivector 
field \cite{lmz}:
\begin{eqnarray}
[[\mathbf{g},\mathbf{h}]]=\sum_{I,J}[v^I,w^J]\otimes[\b v^I,\b w^J].
\end{eqnarray} 
It appears that this bivector field has the explicit expression in terms of $\mathbf{g}$, $\mathbf{h}$ and therefore leads to the following 
definition.

\vspace{3mm} 

\noindent{\bf Definition 2.1.} {\it Let $\mathbf{g},\mathbf{h} \in \Gamma(T'M\otimes T''M)$ written in components as\\ 
$g^{i\b j}\p_i\otimes\p_{\b j}, h^{i\b j}\p_i\otimes\p_{\b j}$.  
Then one can define symmetric bilinear operation 
\begin{eqnarray}
[[,]]:  \Gamma(T'M\otimes T''M)\otimes \Gamma(T'M\otimes T''M)\to \Gamma(T'M\otimes T''M)
\end{eqnarray}
written in components as follows:
\begin{eqnarray}
&&[[\mathbf{g},\mathbf{h}]]^{k\b l}\p_k\otimes\p_{\b l}\equiv\\
&&(g^{i\b j}\p_i\p_{\b j}h^{k\b l}+h^{i\b j}\p_i\p_{\b j}g^{k\b l}-\p_ig^{k\b j}\p_{\b j}h^{i\b l}-
\p_ih^{k\b j}\p_{\b j}g^{i\b l})\p_k\otimes\p_{\b l}.\nonumber
\end{eqnarray}}

\noindent{\bf Property} (Important observation). One can easily find the same structure in Kaehler geometry. If metric associated with $\mathbf{g}$ is 
Kaehler, then the Ricci tensor $R^{i\b j}$, 
expressed in terms of bilinear vector field $g^{i\b j}\p_i\otimes\p_{\b j}$ associated with metric tensor is proportional to 
$[[\mathbf{g},\mathbf{g}]]$, more precisely 
\begin{eqnarray}
R^{i\b j}(\mathbf{g})=\frac{1}{2}[[\mathbf{g},\mathbf{g}]]^{i\b j}.
\end{eqnarray}

\vspace{3mm} 

\noindent
{\bf Definition 2.2.} {\it Suppose the complex manifold $M$ under consideration is equipped with nonvanishing holomorphic top degree form $\Omega$, which in local coordinates can be written as follows:
$\Omega=e^{f(X)}dX^1\wedge...\wedge dX^{D/2}$, where $D$ is the dimension of the manifold. Then one can define:\\ 
1). A covariant divergence $div_{\Omega}: \Gamma (T'M)\to C(M)$ of a section $\mathbf{v}=v^i\p_i\in \Gamma (T'M)$, 
such that 
\begin{eqnarray}
div_{\Omega}\mathbf{v}=\Omega^{-1}\mathcal{L}_\mathbf{v}\Omega=\p_iv^i+\p_i f v^i.
\end{eqnarray}
\\
2). A covariant divergences $div_{\Omega}: \Gamma (T'M\otimes T''M )\to \Gamma (T'M)$ and $div_{\b \Omega}: \Gamma (T'M\otimes T''M )\to \Gamma (T''M)$ of a bivector field such that in local coordinates it has the expression:
\begin{eqnarray}
div_{\Omega}\mathbf{g}=(\p_ig^{i\b j}+\p_ifg^{i\b j})\p_{\b j}\quad 
div_{\b \Omega}\mathbf{g}=(\p_{\b j}g^{i\b j}+\p_{\b j}fg^{i\b j})\p_ i.
\end{eqnarray}
}
We already know from Introduction that the conditions of conformal invariance for the perturbed $\beta$-$\gamma$ 
system is equivalent to the system of Einstein equations with $B$-field and a dilaton. Now we reexpress Einstein equations 
in terms of a bivector field $g^{i\b j}$. 

\vspace{3mm} 

\noindent{\bf Proposition 2.1} 
{\it The equations 
\begin{eqnarray}\label{component}
&&R^{\mu\nu}={1\over 4} H^{\mu\lambda\rho}H^{\nu}_{\lambda\rho}-2\nabla^{\mu}
\nabla^{\nu}\Phi,\nonumber\\
&&\nabla_{\mu}H^{\mu\nu\rho}-2(\nabla_{\lambda}\Phi)H^{\lambda\nu\rho}=0,
\end{eqnarray}
where metric, B-field, and a dilaton are expressed as follows: 
\begin{eqnarray}\label{constr}
G_{i\bar{k}}=g_{i\bar{k}}, \quad B_{i\bar{k}}=-g_{i\bar{k}}, \quad \Phi=\log\sqrt{g}+\Phi_0,
\end{eqnarray}
are equivalent to the following system:
\begin{eqnarray}\label{comp}
&&\p_i\p_{\bar{k}}\Phi_0=0,\quad \p_{\bar{p}}d^{\Phi_0}_{\bar{l}}g^{\bar{l}k}=0, 
\quad \p_{p}d^{\Phi_0}_{l}g^{\bar{k}l}=0,\nonumber\\
&&2g^{r\bar{l}}\p_r\p_{\bar{l}}g^{i\bar{k}}-2\p_r g^{i\bar{p}}\p_{\bar{p}}g^{r\bar{k}}-
g^{i\bar{l}}\p_{\bar{l}}d^{\Phi_0}_sg^{s\bar{k}}-g^{r\bar{k}}\p_r d^{\Phi_0}_{\bar{j}}g^{\bar{j}i}+\nonumber\\
&&\p_rg^{i\bar{k}}d^{\Phi_0}_{\bar{j}}g^{\bar{j}r}+\p_{\bar{p}}g^{\bar{k}i}d^{\Phi_0}_n g^{n\bar{p}}=0,
\end{eqnarray}
where $d^{\Phi_0}_ig^{i\bar{j}}\equiv \p_i g^{i\bar{j}}-2\p_i\Phi_0 g^{i\bar{j}}$ and $d^{\Phi_0}_{\b i}g^{\b i j}\equiv \p_{\b i} g^{j\bar{i}}-2\p_{\b i} \Phi_0 g^{j\bar{i}}$.
}

\vspace{3mm} 

\noindent
{\bf Proposition 2.2} {\it The equation
\begin{eqnarray}\label{dilaton}
4(\nabla_{\mu}\Phi)^2-4\nabla_{\mu}\nabla^{\mu}\Phi+
R+{1\over 12} H_{\mu\nu\rho}H^{\mu\nu\rho}=0,
\end{eqnarray}
where metric, B-field, and dilaton are constrained by \rf{constr} and governed by  equations \rf{component}, is equivalent 
to the following one:
\begin{eqnarray}\label{dil}
d^{\Phi_0}_{i}d^{\Phi_0}_{\bar{j}}g^{\bar{j}i}=0,
\end{eqnarray}
where $d^{\Phi_0}_{i}d^{\Phi_0}_{\bar{j}}g^{\bar{j}i}\equiv (\p_{\b j}-2\p_{\b j}\Phi_0)(\p_i -2\p_i\Phi_0) g^{i\bar{j}}$.
}
\vspace{3mm} 

\noindent
The proofs of the above two propositions are given in Appendix A.\\ 
Before reexpressing the equations on $g^{i\b j}$ 
in terms of the algebraic structures introduced above, we note several interesting properties. 
The first observation is how the equations depend on the bivector field. It is easy to see that 
the dependence is either linear or pure bilinear in $g^{i\b j}$. This is very unusual for the equations of 
Einstein type which are highly nonlinear in the general case.
 
Then one can notice that if there exists a global solution (on the whole manifold $M$) to 
the equations from Proposition 2.1.,
 then the first Chern class $c_1(M)$ should vanish or, in other words, the manifold $M$ 
should possess a holomorphic top form. This is due to one of the equations from the Proposition 2.1., 
namely $\p_i\p_{\b k}\Phi_0=0$. Really, density function
$e^{-2\Phi_0}$ \footnote{Since dilaton $\Phi$ is a well defined function and $det(g_{i\b j})$ is a well 
defined density function for the volume form, therefore $e^{-2\Phi_0}=e^{-2\Phi}det(g_{i\b j})$ is also a density function for the volume form.} 
locally decomposes into the product of holomorphic and antiholomorphic functions $\omega(X)$ and 
$\b \omega(\b X)$ which globally serve as densities for the holomorphic 
and antiholomorphic volume forms $\Omega$ 
and $\b \Omega$, such that    
\begin{eqnarray}
\Omega\b \Omega=e^{-2\Phi_0}dX^1\wedge...\wedge dX^n\wedge dX^{\b 1}...\wedge dX^{\b n}.
\end{eqnarray}
Another property is that $div_{\Omega}(\mathbf{g})=d^{\Phi_0}_ig^{i\bar{j}}\p_{\b j}$ and 
$div_{\b \Omega}(\mathbf{g})=d^{\Phi_0}_{\bar{j}}g^{\bar{j}i}\p_i$ become (again, if the 
Einstein equations from Proposition 2.1. admit the global solution) an antiholomorphic and holomorphic sections of $T''M$ and 
$T'M$ correspondingly. Also, one can easily notice that this condition automatically leads to the equation
\begin{eqnarray}
d^{\Phi_0}_{i}d^{\Phi_0}_{\bar{j}}g^{\bar{j}i}=const.
\end{eqnarray}
This should not be strange since the equation 
\begin{eqnarray}
4(\nabla_{\mu}\Phi)^2-4\nabla_{\mu}\nabla^{\mu}\Phi+
R+{1\over 12} H_{\mu\nu\rho}H^{\mu\nu\rho}=const
\end{eqnarray}
is a direct consequence of equation 
\begin{eqnarray}
\nabla_{\mu}(R^{\mu\nu}-{1\over 4} H^{\mu\lambda\rho}H^{\nu}_{\lambda\rho}+2\nabla^{\mu}
\nabla^{\nu}\Phi)=0.
\end{eqnarray}
So, in our case this is just a particular example of the general statement.
 
Finally, let us summarize and formulate the main result of this section as the following Proposition. 
 
\vspace{3mm} 

\noindent
{\bf Proposition 2.3.} {\it The system of equations 
\rf{component} and \rf{dilaton} with additional constraint \rf{constr} on the manifold $M$ 
is equivalent to the following}:\\
1). {\it There exist a nonvanishing holomorphic top degree form $\Omega=\rho(X)dX^1\wedge...\wedge dX^{D/2}$
on the manifold M (and therefore the first Chern class $c_1(M)$ should vanish) such that}  
\begin{eqnarray}
\rho(X)\b \rho(\b X)=e^{-2\Phi(X, \b X)}g(X, \b X).  
\end{eqnarray}
2). {\it Vector fields $div_{\b \Omega}(\mathbf{g})\in \Gamma (T'M)$ and  $div_{\Omega}(\mathbf{g})\in \Gamma(T''M)$ are respectively 
holomorphic and antiholomorphic.}\\
3). {\it Bivector field $\mathbf{g}\in \Gamma(T'M\otimes T''M)$ obeys the following two equations: 
\begin{eqnarray}\label{bil}
[[\mathbf{g},\mathbf{g}]]+\mathcal{L}_{div_{\Omega}(\mathbf{g})}\mathbf{g}+\mathcal{L}_{div_{\b \Omega}(\mathbf{g})}\mathbf{g}=0, 
\quad div_{\b \Omega}div_{\Omega}(\mathbf{g})=0,
\end{eqnarray} 
where $\mathcal{L}_{div_{\Omega}(\mathbf{g})}$ and $\mathcal{L}_{div_{\b \Omega}(\mathbf{g})}$ are Lie derivatives with respect to 
the corresponding vector fields.}

\vspace{3mm} 

\noindent
{\bf Remark.} It's easy to see that when $g_{i\b j}$ is the Kaehler metric, 
then the bilinear equation on $\bf g$ from \rf{bil} is automatically satisfied and the equations \rf{component}, \rf{dilaton} with constraint 
\rf{constr} reduce to linear ones.

\section{Perturbed $\beta$-$\gamma$ System and Generalized \\
Maurer-Cartan Equations}

\subsection{Motivation}

From  section 2 we learned that the conditions of conformal invariance for $\beta$-$\gamma$ system perturbed by 
$g^{i\b j}p_ip_{\b j}$ operator are given by  bilinear and linear equations on the corresponding bivector field. 
In paper \cite{lmz} the conditions of conformal invariance for 
this first order sigma model were studied naively via cut-off regularization. The resulting equations
 which correspond 
to the case when $\Phi_0=0$ (the additional dilaton field was ignored there) 
and $g^{i\b j}p_ip_{\b j}$ is a primary field ($\p_ig^{i\b j}=0$ and $\p_{\b j}g^{i\b j}=0$) 
have the following form:
\begin{eqnarray}
[[\mathbf{g},\mathbf{g}]]=0.
\end{eqnarray}
Introducing the standard ghost fields $b$, $c$  and $\t b$, $\t c$ with operator products 
$c(z)b(w)\sim \frac{1}{z-w}$ and $\t c(z)\t b(w)\sim \frac{1}{\b z-\b w}$, and BRST operator \cite{brst}, \cite{pol}:
\begin{eqnarray}
\label{BRST}
&&Q=\frac{1}{2\pi i}\oint\mathcal{J_B},\quad \mathcal{J_B}=j_Bdz-\tilde{j_B}d\bar{z},\\
&&j_B=cT+:bc\partial c:, \quad \tilde{j}_B=\tilde{c}\tilde{T}
+:\tilde{b}\tilde{c}
\bar{\partial} \tilde{c}:,\nonumber
\end{eqnarray}
where $T$, $\t T$ are the holomorphic and antiholomorphic components of the energy-momentum tensor,
we found out that the resulting equations can be rewritten as follows:
\begin{eqnarray}\label{sen}
[Q,\psi^{(0)}]=0, \quad \lim_{h\to 0}\mathcal{P}\int_{C_{\epsilon,z}}\psi^{(1)}\psi^{(0)}(z)=0,
\end{eqnarray}
where $\psi^{(0)}=\t c ch^{-1}g^{i\b j}p_ip_{\b j}$, $\psi^{(1)}=dz \t c h^{-1}g^{i\b j}p_ip_{\b j}- 
d\b z c h^{-1}g^{i\b j}p_ip_{\b j}$, $C_{\epsilon,z}$ is the contour around point $z$, and $\mathcal{P}$ 
is the projection on the $\epsilon$-independent terms\footnote{Here we 
point out that the bilinear operation similar to \rf{sen} 
was considered in \cite{sen} in the context very close to ours, namely the comparison of the conditions of BRST-invariance of the conformal perturbation 
theory and the string field theory equations of motion.}.\\ 
This suggests that the complete system of equations given in Proposition 2.3. and also others entering beta function with higher orders in 
$h$ should be written in the Maurer-Cartan form:
\begin{eqnarray}
[Q,\phi^{(0)}]+M(\phi^{(0)},\phi^{(0)})+...=0,
\end{eqnarray}
where 
\begin{eqnarray}
M(\phi^{(0)},\phi^{(0)})(z)\sim \mathcal{P}\int_{C_{\epsilon,z}}\phi^{(1)}\phi^{(0)}(z)
\end{eqnarray}
is a bilinear operation and $\phi^{(0)}$, $\phi^{(1)}$ are some modifications of $\psi^{(0)}$, $\psi^{(1)}$. 
In  subsection 3.2 we define properly $\phi^{(0)}$, $\phi^{(1)}$, and $M$.

\subsection{Maurer-Cartan Form of Conformal Invariance Conditions.}

{\bf 1. Notation and conventions.}

\vspace{3mm} 

\noindent  
{\bf Descent hierarchy and ghost fields.} Throughout this section we will denote a conformal field as a 0-form with the corresponding superscript if it is a 
differential polynomial in $c, \t c$ ghost fields, where the matter fields are coefficients, and denote the space of such fields as 
$H^0$. 
We also introduce the ghost number operator
\begin{eqnarray}
N_g=\int (dzj_g-d\b z \t j_g),  
\end{eqnarray}
where $j_g=-bc$ and $\t j_g=-\t b \t c$. If $\phi^{(0)}$ is the eigenvector of this operator with the eigenvalue $n_{\phi}$, we 
say that this field is of ghost number $n_{\phi}$ (it is obvious that it can be only nonnegative integer).  
We associate with any field $\phi^{(0)}\in H^0$ depending on matter and ghost $c, \t c$  fields the 
following 1-form and 2-form: 
\begin{eqnarray}
\phi^{(1)}=dz[b_{-1},\phi^{0}]+d\b z[\t b_{-1},\phi^{(0)}], \quad
\phi^{(2)}=dz\wedge d\b z[b_{-1},[\t b_{-1},\phi^{(0)}]].  
\end{eqnarray}
{\bf Operator products.} We assume that all matter field operators enjoy 
the operator products of the following type:
\begin{eqnarray}\label{ope}
V(z)W(z')=\sum_{r=-\infty}^{m}\sum_{s=-\infty}^{n}(V,W)^{(r,s)}(z')(z-z')^{-r}(\b z-\b z')^{-s}.
\end{eqnarray} 
Throughout this section we assume that all operators are ordered with respect to holomorphic normal ordering. We neglect the sign of 
normal ordering if it does not lead to misunderstanding.

\vspace{3mm} 

\noindent
{\bf 2. Bilinear operation and its properties.} First of all we give the definition of
 the bilinear operation itself.

\vspace{3mm} 

\noindent
{\bf Definition 3.1.} {\it For any two fields $\phi^{(0)}$, $\psi^{(0)}$ we define the bilinear operation
$M:H^0 \otimes H^0\to H^0$
\begin{eqnarray}
&&M(\phi^{(0)}, \psi^{(0)})(z)=\\
&&\frac{1}{4\pi i}\mathcal{P}\int_{C_{\epsilon,z}}\phi^{(1)}\psi^{(0)}(z)+(-1)^{n_{\phi}n_{\psi}}
\frac{1}{4\pi i}\mathcal{P}\int_{C_{\epsilon,z}}\psi^{(1)}\phi^{(0)}(z),\nonumber
\end{eqnarray}
where $\mathcal{P}$ is a projection on the $\epsilon^0$ term.  }

\vspace{3mm} 

\noindent
{\bf Remark} (Comment about the projection operator). Really, 
for any 
$\chi_1^{(0)}$, $\chi_2^{(0)}$ under consideration 
$\int_{C_{\epsilon,z}}\chi_1^{(1)}\chi_2^{(0)}(z)$ lies in the space of formal power series 
$H^{0}(\epsilon)=\{\phi_{\epsilon}^{(0)}$ $|$\ $\phi_{\epsilon}^{(0)}=\sum^{\infty}_{n=-k}\epsilon^n\phi_{n}^{(0)}$, 
$\phi_{n}^{(0)}\in H^0\}$. 
Therefore, the operator $\mathcal{P}$ is well defined and projects on the coefficient of $\epsilon^0$ of the corresponding element 
of $H^{0}(\epsilon)$. 

\vspace{3mm} 

\noindent
Definition 3.1. leads to the following properties.

\vspace{3mm} 

\noindent
{\bf Proposition 3.1.}{\it Operation $M$ satisfies the following relation:
\begin{eqnarray}\label{2prod}
[Q,M(\phi^{(0)}, \psi^{(0)})]+M( [Q,\phi^{(0)}], \psi^{(0)})+(-1)^{n_{\phi}}M( \phi^{(0)}, [Q,\psi^{(0)}])=0.
\end{eqnarray}}
{\bf Proof.} First, we need to show that BRST operator commutes with projection operator $\mathcal{P}$. Really, let's denote 
$
f(V,W)(z)=\int_{C_{\epsilon,z}}dwV(w)W(z)
$ 
for some operators $V$, $W$. 

From \rf{ope} we know that $f(V,W)=\sum^{\infty}_{n=-k}f_n(V,W)\epsilon^n$. The projection operator acts as follows:
$\mathcal{P}f(V,W)=f_0(V,W)$. Therefore we see that $\mathcal{P}[Q,f(V,W)]=[Q,\mathcal{P}f(V,W)]=[Q,f_0(V,W)]$. 
In such a way we see that BRST operator commutes with projection operator and hence the relation \rf{2prod} 
can be easily established by means of the simple formula 
$[Q, \phi^{(1)}]=d\phi^{(0)}-[Q, \phi^{(0}]^{(1)}$. $\blacksquare$

\vspace{3mm} 

\noindent
{\bf Remark.} Relation 
 \rf{2prod} is similar to the basic property of the string 2-products \cite{zwiebach}.

\vspace{3mm} 

\noindent
{\bf Proposition 3.2.} {\it The expression for the 2-form associated with $M(\phi^{(0)}, \psi^{(0)})$ is given 
by the following formula:
\begin{eqnarray}
&&M(\phi^{(0)}, \psi^{(0)})^{(2)}=\frac{1}{2\pi i}\mathcal{P}\int_{C_{\epsilon,z}}\phi^{(1)}\psi^{(2)}(z)+\nonumber\\
&&(-1)^{n_{\phi}n_{\psi}}\frac{1}{2\pi i}\mathcal{P}\int_{C_{\epsilon,z}}\psi^{(1)}\phi^{(2)}(z)+d\chi^{(1)},
\end{eqnarray}
where as usual $M(\phi^{(0)}, \psi^{(0)})^{(2)}=dz\wedge d\b z[b_{-1},[\t b_{-1},M(\phi^{(0)}, \psi^{(0)})]]$, 
 $\chi^{(1)}$ is some 1-form, and $d$ is the de Rham differential.}

\vspace{3mm} 

\noindent
The proof can be easily obtained from Proposition 2.1. of \cite{zeit}.

\vspace{3mm} 

\noindent
{\bf 3. Maurer-Cartan Structures and $\beta$-$\gamma$ Systems.} 
We suggest (according to the hypothesis of A.S. Losev \cite{losev}) that the equations governing the conformal invariance of  system \rf{bgp} can be summarized via the following generalized Maurer-Cartan equation:
\begin{eqnarray}\label{gmc}
C_1(\phi^{(0)})+C_2(\phi^{(0)},\phi^{(0)})+C_3(\phi^{(0)},\phi^{(0)},\phi^{(0)} )+...=0,
\end{eqnarray}
such that $C_n$ are graded (with respect to ghost number) symmetric bilinear operations, 
$C_1(\phi^{(0)})=[Q_{\omega}, \phi^{(0)}]$ and $C_2=\frac{1}{2}M$, where $Q_{\omega}$ is the BRST operator from 
the previous section such that holomorphic and the antiholomorphic components of the energy-momentum tensor 
are those for accurately defined $\beta$-$\gamma$ system (see section 1):
\begin{eqnarray}
&&Q_{\omega}=\int \mathcal{J}_{\omega},\\ 
&&\mathcal{J}_{\omega}=dzc(-h^{-1}p_i\p X^i-1/2\p^2\log \omega)-d\b z\t c(-h^{-1}p_{\b i}\bp X^{\b i}-1/2\bp^2\log \b \omega),\nonumber
\end{eqnarray}
the 2-form $\phi^{(2)}$ associated with $\phi^{(0)}$ is the ``perturbation'' 2-form
$\phi^{(2)}=dz\wedge d\b z V$ such that $V=h^{-1}g^{i\b j}p_ip_{\b j}$. We also put the following constraints on $\phi^{(0)}$: 
\begin{eqnarray}\label{semi}
b^{-}_0\phi^{(0)}=0, 
\end{eqnarray}
where $b^{-}_0=b_0-\t b_0$ .\\
This gives the following expression for $\phi^{(0)}$:
\begin{eqnarray}
\phi^{(0)}=\t cc V+c(\p c +\bp \t c)W- \t c  (\p c +\bp \t c)\b W+1/2c\p^2 c U-1/2\t c\bp^2 \t c\b U -\nonumber\\
\t c\p^2 cX+c\bp^2 \t c \b X-(\p c+\bp \t c)\p^2 c Y+(\p c+\bp \t c)\bp^2 c \b Y+\dots,
\end{eqnarray}
where $W$, $\b W$, $U$, $\b U$, $X$, $\b X$, $Y$, $\b Y$ are some matter fields and $\dots$ stand for the terms depending on $\p^nc$ or $\bp^m\t c$ such that $n,m>2$. 
We will refer to the terms with $U$ and $\b U$ as $dilatonic$ terms. Moreover we make these dilatonic terms only $X$, $\b X$-dependent, that is $U=f(X,\b X)$, $\b U=\b f(X,\b X)$. The $W$, $\b W$ terms will be called in the following as $gauge$ terms, so we will refer to the 
constraint $b_0\phi^{(0)}=\t b_0\phi^{(0)}=0$ (which is equivalent to $W=\b W=0$) as a $gauge$ $condition$. 
The $X$, $\b X$, $Y$, $\b Y$-terms have no physical interpretation, therefore we get rid of them reducing to the subspace.

\vspace{3mm} 

\noindent
{\bf Definition 3.2.} {\it The space $S^0$ consists of the elements $\phi^{(0)}\in H^0$ which enjoy three properties}:\\
{\bf 1.} $n_{\phi}=2$,\\
{\bf 2.} $b^{-}_0\phi^{(0)}=0$,\\
{\bf 3.} $b_{i}\t b_{j}\phi^{(0)}=0$ {\it if $i+j>-1$}.

\vspace{3mm} 

\noindent
{\bf Remark.} As we see the general form of the element from $\phi^{(0)}\in S^0$ is as follows:
\begin{eqnarray}
\phi^{(0)}=\t cc V+c(\p c +\bp \t c)W- \t c  (\p c +\bp \t c)\b W+1/2c\p^2 c U-1/2\t c\bp^2 \t c\b U +\dots,
\end{eqnarray}
where $\dots$ stand for the terms depending on $\p^nc$ or $\bp^m\t c$ such that $n,m>2$.

\vspace{3mm} 

\noindent
If we expand the field $\phi^{(0)}$ in the series of some formal parameter $t$: $\phi^{(0)}=\sum^{\infty}_{n=1}\phi_n^{(0)}t^n$, then the first 
two orders of expansion of  equation \rf{gmc} give the following equations:
\begin{eqnarray}\label{mc}
[Q_{\omega}, \phi_1^{(0)}]=0, \quad  [Q_{\omega}, \phi_2^{(0)}]+\frac{1}{2}M(\phi_1^{(0)},\phi_1^{(0)}) =0.
\end{eqnarray}
We show that \rf{mc} lead to the equations \rf{comp} and \rf{dil} expanded to the second order in the formal parameter. 
Let's formulate this as proposition.

\vspace{3mm} 

\noindent
{\bf Proposition 3.3.} {\it Let $\phi^{(0)}$ be the element of $S^0$, satisfying the following conditions:
$\phi^{(2)}=dz\wedge d\b z h^{-1}g^{i\b j}(X, \b X)p_ip_{\b j}$, the dilatonic terms $U(X,\b X)=b_{1}b_{-1}\phi^{(0)}$, and 
$\b U(X, \b X)=\t b_{1}\t b_{-1}\phi^{(0)}$ are respectively functions of $X^i(z)$,  $X^{\b i}(\b z)$. Then \rf{mc}  
gives the equations \rf{comp} and \rf{dil} expanded up to the second order in the formal parameter t, 
such that the expansion of matter fields is: $g^{i\b j}=tg_1^{i\b j}+t^2g_2^{i\b j}+...$, 
$\Phi_0=1/2(-\log(\omega\b \omega)+(tU_1+t\b U_1)+t^2(U_2+\b U_2)+...)$.}

\vspace{3mm} 

\noindent
The Proof is given in  Appendix 2.

\vspace{3mm} 

\noindent
{\bf Remark 1.} One could also put the condition that $U, \b U$ are holomorphic and antiholomorphic functions
correspondingly. 
Really, we can consider $Q+\phi^{(1)}$ as a deformation of a BRST current \cite{verlinde},\cite{zeit}. From the proof of Proposition 3.3. one can 
see that adding a total derivative to $\phi^{(1)}$ we can get rid of 
non(anti)holomorphic terms from $\b U_1$ ($U_1$). Hence it is natural to put this additional constraint.

\vspace{3mm} 

\noindent
{\bf Remark 2.} Applying $b_{-1}\t b_{-1}$ to \rf{mc} and using the Proposition 3.2., we get the following equations:
\begin{eqnarray}
[Q_{\omega}, \phi_1^{(2)}]=d\phi_1^{(1)}, \quad [Q_{\omega}, \phi_2^{(2)}](z)+
\frac{1}{2\pi i}\mathcal{P}\int_{C_{\epsilon,z}}\phi_1^{(1)}\phi_1^{(2)}(z) =d\psi_2^{(1)}(z)
\end{eqnarray}
which can be interpreted as a conservation law for the deformed BRST current in the presence of perturbation $\phi^{(2)}$ \cite{zeit}.

\vspace{3mm} 

\noindent
{\bf Remark 3.} From the Proposition 3.1. we know that there are relations between operations $C_1$ and $C_2$ from 
\rf{gmc}:
\begin{eqnarray}
&&C_1(C_1(\phi^{(0)}))=0, \\ 
&&C_1(C_2(\phi^{(0)}, \psi^{(0)}))+C_2( C_1(\phi^{(0)}), \psi^{(0)})+(-1)^{n_{\phi}}C_2( \phi^{(0)}, C_1(\psi^{(0)}))=0\nonumber
\end{eqnarray}
for any $\phi^{(0)},\psi^{(0)}\in H^0$. We hope that similar quadratic relations will hold for all $C_n$ and generate the homotopy Lie algebra \cite{zwiebach}, \cite{stasheff}, like it was for string products. 

\vspace{3mm} 

\noindent
{\bf 4. Symmetries of Maurer-Cartan Equation.} In paragraph 3. we made a conjecture that the conditions of conformal invariance 
for  model \rf{bgp} coincide with a sort of 
Maurer-Cartan equation. 
However, we want our equation to possess symmetries, more precisely we want them to be in the following form:
\begin{eqnarray}
\delta\phi^{(0)}=\varepsilon(C_1(\xi^{(0)})+C_2(\phi^{(0)},\xi^{(0)})+C_3(\phi^{(0)},\phi^{(0)},\xi^{(0)} )+...),
\end{eqnarray} 
where $\varepsilon$ is infinitesimal and $\xi^{(0)}\in H^0$ is of ghost number 1. We will see now that at the lowest order in the parameter $h$ 
they have precisely this form. Really, let's look on  equations \rf{mc}. 
The first equation of \rf{mc} due to the nilpotence of operator $Q_{\omega}$ has the following symmetry:
\begin{eqnarray}
\delta\phi_1^{(0)}=\varepsilon[Q_{\omega},\xi_1^{(0)}],
\end{eqnarray}
where $\xi_1^{(0)}$ is a zero form of ghost number 1 and $\varepsilon$ is infinitesimal. 
Proposition 3.1. allows to accompany this with the following transformation:
\begin{eqnarray}
\delta\phi_2^{(0)}=\varepsilon([Q_{\omega},\xi_2^{(0)}]+ M(\xi_1^{(0)},\phi_1^{(0)})),
\end{eqnarray}
where $n_{\xi_2}=1$ is again a 0-form of ghost number 1. Altogether they form a symmetry of \rf{mc}.
In our case these  symmetry transformations should correspond to the symmetries of \rf{comp}, \rf{dil}, i.e. holomorphic coordinate transformations, therefore it is natural to give the following expression for $\xi^{(0)}=\sum^{\infty}_{n=1}t^n\xi_n^{(0)}$:
\begin{eqnarray}\label{xi}
 \xi^{(0)}=h^{-1}(cv^i(X)p_i-\tc \b v^{\b i}(\b X)p_{\b i}), 
\end{eqnarray}
 where $v^i(X)$ are the components of the holomorphic section of $T'M$. 
It appears that the second order approximation to symmetries considered above is very close to 
exact expression in our case.

\vspace{3mm} 

\noindent
{\bf Proposition 3.4.} {\it 
Let $\phi^{(0)}\in S^0$ satisfy the following conditions:
$\phi^{(2)}=dz\wedge d\b z h^{-1}g^{i\b j}(X, \b X)p_ip_{\b j}$,  the dilatonic terms be $U(X, \b X)=b_{1}b_{-1}\phi^{(0)}$, and 
$\b U(X,\b X)=\t b_{1}\t b_{-1}\phi^{(0)}$. Then 
\begin{eqnarray}\label{sym}
\delta_{\xi}\phi^{(0)}=\varepsilon([Q_{\omega},\xi^{(0)}]+M(\xi^{(0)},\phi^{(0)})),
\end{eqnarray}
where $\xi^{(0)}$ is given by \rf{xi} and $\varepsilon$ is infinitesimal, gives the following transformations:
\begin{eqnarray}
&&\delta_{\xi} \phi^{(2)}=dz\wedge d\b z h^{-1}(\delta_{\bf v}g^{i\b j}p_ip_{\b j}+O(h)),\nonumber\\ 
&&\delta_{\xi} \Phi_0(X,\b X)=\varepsilon (\half div_{\Omega}{\bf v}(X)+\half div_{\b \Omega}{\bf \b v}(\b X)),
\end{eqnarray}
where $ \delta_{\bf v} {\bf g}=-\varepsilon(L_{\bf v}{\bf g}+L_{\bf \b v}{\bf g})$ which coincides with the holomorphic coordinate transformation of 
{\bf g} and $\Phi_0=1/2(-\log{(\omega\b \omega)}+U+\b U)$.}\\
The proof can be obtained easily by the direct calculation.

\vspace{3mm} 

\noindent
{\bf Remark.} The appearance of the additional (noncovariant) terms of the higher order in $h$
 in $\phi^{(2)}$ after  symmetry 
transformation \rf{sym} has deep roots in the very 
nature of curved $\beta$-$\gamma$ systems (see e.g. \cite{schekhtman}, \cite{nekrasov}). We will consider them elsewhere.

\subsection{Relations of $M$-operation with Courant and Dorfman Brackets.} 

In this subsection we discuss a simple property of bilinear operation $M$ which may 
lead to  some deep consequences in the understanding of the theory.

At first we remind the definition and point out some useful properties of Courant \cite{courant} and Dorfman \cite{dorfman} brackets which appear to be very important in the 
theory of generalized complex structures 
\cite{hitchin}, \cite{gualtieri}.

\vspace{3mm} 

\noindent
{\bf Definition 3.3.} {\it Let $({\bf v}_i,{\bf w}_i)={\bf v}_i+{\bf w}_i$ (i=1,2) be the sections of $TM\oplus T^*M$. Courant and Dorfman brackets 
$[$ , $]_c$, $[$ , $]_d$: $\Gamma(TM\oplus T^*M)\otimes \Gamma(TM\oplus T^*M) \to \Gamma(TM\oplus T^*M$) are defined as follows:
\begin{eqnarray}
&&[ ({\bf v}_1,{\bf w}_1), ({\bf v}_2,{\bf w}_2)]_c=([{\bf v}_1,{\bf v}_2],L_{{\bf v}_1}{\bf w}_2-L_{{\bf v}_2}{\bf w}_1-
\frac{1}{2}d(\mathbf{i}_{{\bf v}_1}{\bf w}_2-\mathbf{i}_{{\bf v}_2}{\bf w}_1)),\nonumber\\
&&[ ({\bf v}_1,{\bf w}_1), ({\bf v}_2,{\bf w}_2)]_d=([{\bf v}_1,{\bf v}_2],L_{{\bf v}_1}{\bf w}_2-\mathbf{i}_{{\bf v}_2}d{\bf w}_1).
\end{eqnarray}}
Now we give the properties which provide the differences between Courant and Dorfman brackets and the Lie one.

\vspace{3mm} 

\noindent
{\bf Properties.}\\
1. {\it The Dorfman bracket satisfies the Leibnitz rule, but it is not antisymmetric, moreover its antisymmetrization leads to the Courant bracket, namely:
\begin{eqnarray}
&&[ ({\bf v}_1,{\bf w}_1), ({\bf v}_2,{\bf w}_2)]_d+[ ({\bf v}_2,{\bf w}_2), ({\bf v}_1,{\bf w}_1)]_d=
d(({\bf v}_1,{\bf w}_1), ({\bf v}_2,{\bf w}_2)),
\\
&&\frac{1}{2}([ ({\bf v}_1,{\bf w}_1), ({\bf v}_2,{\bf w}_2)]_d-
[ ({\bf v}_2,{\bf w}_2), ({\bf v}_1,{\bf w}_1)]_d)=[ ({\bf v}_2,{\bf w}_2), ({\bf v}_1,{\bf w}_1)]_c,\nonumber
\end{eqnarray}
where $(({\bf v}_1,{\bf w}_1), ({\bf v}_2,{\bf w}_2)=v_1^{\mu}{w_2}_{\mu}+v_2^{\mu}{w_{1}}_{\mu}$ is the symmetric bilinear form on $\Gamma(TM\oplus 
T^*M)$ and $d$ is the de Rham differential.}\\
2. {\it The Courant bracket is antisymmetric but it does not satisfy Jacobi identity, more precisely it satisfies Jacoby identity modulo exact 
(with respect to de Rham differential) term:  
\begin{eqnarray} \label{jc}
&&[({\bf v}_1,{\bf w}_1), [({\bf v}_2,{\bf w}_2), ({\bf v}_3,{\bf w}_3)]_c]_c+
[({\bf v}_3,{\bf w}_3), [({\bf v}_1,{\bf w}_1), ({\bf v}_2,{\bf w}_2)]_c]_c+\nonumber\\
&&[({\bf v}_2,{\bf w}_2), [({\bf v}_3,{\bf w}_3), ({\bf v}_1,{\bf w}_1)]_c]_c=d\mathcal{N}(({\bf v}_1,{\bf w}_1), ({\bf v}_2,{\bf w}_2), ({\bf v}_3,{\bf w}_3)),
\end{eqnarray}
where $\mathcal{N}$ is the Nijenhuis operator} \cite{gualtieri}.

\vspace{3mm} 

\noindent
For the proof and much other information on the subject including references see e.g. \cite{gualtieri}.\\
Recall that in  paragraph 4. of subsection 3.2. 
we studied the symmetries of the generalized Maurer-Cartan equation \rf{mc}. The symmetries were related to the 
0-form $\xi^{(0)}=h^{-1}(cv^i(X)p_i-\tc \b v^{\b i}(\b X)p_{\b i})$, where $v^i(X)$ are the components of the holomorphic section of $T'M$. 
Let's extend this 0-form to the following one:
\begin{eqnarray} 
\xi_{\bf v,\bf w}^{(0)}=h^{-1}c(v^i(X)p_i-w_k(X)\p X^k)-h^{-1}\tc (\b v^{\b i}p_{\b i}-\b w_k(\b X)\p X^k)
\end{eqnarray}
associated with a holomorphic section $({\bf v},{\bf w})=v^i(X)\p_i+w_k(X)d X^k\in \Gamma(T'M\oplus T'M^*)$ \footnote 
{We expect that $\xi_{\bf v,\bf w}^{(0)}$ will generate holomorphic symmetries (not only coordinate invariance symmetry but also the one associated with B-field) of the general first order sigma model, see \cite{lmz} and Conclusions of this paper.}. Let's denote 
also $M_0(\phi^{(0)},\psi^{(0)})=
\lim_{h\to 0}hM(\phi^{(0)},\psi^{(0)})$ if such a limit exists. 

\vspace{3mm} 

\noindent
{\bf Proposition 3.5.} {\it For any two holomorphic sections $({\bf v}_i,{\bf w}_i)={\bf v}_i+{\bf w}_{i}\in \Gamma(T'M\oplus T'M^*)$  
$(i=1,2)$ 
\begin{eqnarray} 
M_0(\xi_{{\bf v}_1,{\bf w}_1}^{(0)},\xi_{{\bf v}_2,{\bf w}_2}^{(0)})=-h\xi_{[({\bf v}_1,{\bf w}_1),({\bf v}_2,{\bf w}_2)]_c}^{(0)}.
\end{eqnarray}}\\
The proof can be obtained by straightforward calculation. 

This correspondence between $M_0$ and Courant bracket leads to the following proposition.

\vspace{3mm} 

\noindent
{\bf Proposition 3.6.} {\it Let $\xi^{(0)}_i=\xi_{{\bf v}_i,{\bf w}_i}^{(0)}$ $(i=1,2,3)$ be the 0-forms associated with 
holomorphic sections $({\bf v}_i,{\bf w}_i)\in  T'M\oplus T'M^*$. Then we have:
\begin{eqnarray}
&&M_0(\xi_1^{(0)}, M(\xi_2^{(0)},\xi_3^{(0)}))+M_0(\xi_3^{(0)}, M(\xi_1^{(0)},\xi_2^{(0)}))\nonumber\\
&&+ M_0(\xi_2^{(0)}, M(\xi_3^{(0)},\xi_1^{(0)}))+[Q_{\omega},\mathcal{N}_{1,2,3}(X)+\b \mathcal{N}_{1,2,3}(\b X)]=0,
\end{eqnarray}
where $\mathcal{N}_{1,2,3}(X)=\mathcal{N}(({\bf v}_1,{\bf w}_1), ({\bf v}_2,{\bf w}_2), ({\bf v}_3,{\bf w}_3))(X)$ and $\b \mathcal{N}_{1,2,3}(\b X)$ is its 
complex conjugate.}\\
{\bf Proof.} The proof is the immediate consequence of Proposition 3.5., property \rf{jc} of the Courant bracket, 
and the fact that $[Q_\omega, N_{1,2,3}+\b N_{1,2,3}]=c\p{X^i}\p_iN_{1,2,3}+\t c \bp{X^{\b i}}\p_{\b i}\b N_{1,2,3}$. 
$\blacksquare$ 

\vspace{3mm} 

\noindent 
{\bf Remark.}We see that  operation $M$ acting on the space of the zero forms $\xi_{{\bf v},{\bf w}}^{(0)}$ 
is antisymmetric and in the classical limit $h\to 0$ obeys Jacobi identity modulo $Q_{\omega}$-exact terms. This property reminds the one unifying string 
2- and 3-products leading to $L_{\infty}$-algebra \cite{zwiebach}. Therefore if we 
identify the operation $M$ with 2-product, then 3-product for three $\xi_{{\bf v}_i,{\bf w}_i}^{(0)}$ 0-forms in the classical limit 
gives Nijenhuis operator acting on the corresponding sections of $TM\oplus T^*M$. 

\vspace{3mm}

\noindent 
From the definition of $M$ it is evident that this operation is the supersymmetrization of the following one:
\begin{eqnarray}
N(\phi^{(0)}, \psi^{(0)})=\frac{1}{2\pi i}\mathcal{P}\int_{C_{\epsilon,z}}\phi^{(1)}\psi^{(0)},
\end{eqnarray}
that is $M(\phi^{(0)}, \psi^{(0)})=1/2(N(\phi^{(0)}, \psi^{(0)})+(-1)^{n_{\phi}n_{\psi}}N(\psi^{(0)}, \phi^{(0)}))$. 
Therefore it is reasonable to think about the correspondence of $N_0=\lim_{h\to 0}hN$ with the Dorfman bracket, in the sence of the correspondence between 
$M_0$ and the Courant one.

\vspace{3mm} 

\noindent
{\bf Proposition 3.7.} {\it Let's take $\xi_{{\bf v}_1,{\bf w}_1}^{(0)}$} $(i=1,2)$ {\it as in Proposition 3.5. Then the following holds:}\\
\begin{eqnarray}
&&\label{ndorf}N_0(\xi_{{\bf v}_1,{\bf w}_1}^{(0)},\xi_{{\bf v}_2,{\bf w}_2}^{(0)})=-h\xi^{(0)}_{[({\bf v}_1,{\bf w}_1),({\bf v}_2,{\bf w}_2)]_d},\\
&&[Q_{\omega},f_{12}(X)+\b f_{12}(\b X)] =
N_0(\xi_{{\bf v}_1,{\bf w}_1}^{(0)},\xi_{{\bf v}_2,{\bf w}_2}^{(0)}))+
N_0(\xi_{{\bf v}_2,{\bf w}_2}^{(0)},\xi_{{\bf v}_1,{\bf w}_1}^{(0)}),\nonumber
\end{eqnarray}
{\it where $f_{12}(X)=(({\bf v}_1,{\bf w}_1),({\bf v}_2,{\bf w}_2))(X)$ and $\b f_{12}(\b X)$ is its complex conjugate.}

\vspace{3mm} 

\noindent
{\bf Remark 1.} It is worth noting that relation \rf{ndorf} gives a very easy proof of the Leibnitz rule for Dorfman bracket. The proof follows 
directly from the corresponding vertex operator algebra axiom.  

\vspace{3mm} 

\noindent
{\bf Remark 2.} Similar statements to two propositions above were given in \cite{alekseev} in the context of anomalous Poisson brackets in the first 
order theories.\\
The propositions 3.6 and 3.7. are the consequences of the general fact below.

\vspace{3mm} 

\noindent
{\bf Proposition 3.8.} {\it Let $\chi_{i}^{(0)}=cJ_i(z)dz-\t c\b J_i(\b z)d\b z$ (i=1,2,3 ), where $J_i(z)$ ($\b J_i(\b z)$) are (anti) holomorphic 
matter field of conformal weight (1,0) ((0,1)). Then}\\ 
1.{\it The N-operation is antisymmetric modulo BRST-exact term:
\begin{eqnarray}
N(\chi_{1}^{(0)},\chi_{2}^{(0)})+N(\chi_{2}^{(0)},\chi_{1}^{(0)})=[Q,\nu_{12}^{(0)}],
\end{eqnarray}
where $\nu^{(0)}_{12}$ is some 0-form of ghost number 0, and satisfies Leibnitz rule on $\chi_{i}^{(0)}$.}\\
2.{\it The M-operation which is supersymmetrization of the $N$-operation, satisfies Jacobi identity for the fields of the form  $\chi^{(0)}$ modulo 
BRST-exact term:
\begin{eqnarray}\label{jac}
&&M(\chi_1^{(0)}, M(\chi_2^{(0)},\chi_3^{(0)}))+M(\chi_3^{(0)}, M(\chi_1^{(0)},\chi_2^{(0)}))\\
&&+ M(\chi_2^{(0)}, M(\chi_3^{(0)},\chi_1^{(0)}))=[Q,\mu^{(0)}_{123}],\nonumber
\end{eqnarray}
where $\mu^{(0)}_{123}$ is some 0-form of ghost number 0.}\\
{\bf Proof.} In order to prove the first part of the proposition one needs to consider the operator product of $J_i(z_1)J_j(z_2)$:
\begin{eqnarray}
&&J_i(z_1)J_j(z_2)\sim\frac{(J_i,J_j)^{(2,0)}(z_2)}{(z_1-z_2)^2}+\frac{(J_i,J_j)^{(1,0)}(z_2)}{(z_1-z_2)},\nonumber\\
&&\b J_i(z_1)\b J_j(z_2)\sim\frac{(\b J_i,\b J_j)^{(0,2)}(z_2)}{(z_1-z_2)^2}+\frac{(\b J_i,\b J_j)^{(0,1)}(z_2)}{(z_1-z_2)}.
\end{eqnarray}
Here $(J_i,J_j)^{(2,0)}$  and $(\b J_i,\b J_j)^{(0,2)}$ are the conformal fields of dimension $(0,0)$. They lead to the relations:
\begin{eqnarray}
&&Res_{z_1\to z_2} (J_i(z_1)c(z_2)J_j(z_2))+Res_{z_1\to z_2} (J_j(z_1)c(z_2)J_i(z_2))=\nonumber\\
&&c(z_2)L_{-1}(J_i,J_j)^{(2,0)}(z_2),\nonumber\\
&&Res_{z_1\to z_2} (\b J_i(\b z_1)\t c(\b z_2)\b J_j(\b z_2))+Res_{z_1\to z_2} (\b J_j(z_1)\t c(z_2)\b J_i(\b z_2))=\nonumber\\
&&\t c(z_2)\b L_{-1}(\b J_i,\b J_j)^{(0,2)}(z_2).
\end{eqnarray}
This immediately leads to the first statement. Moreover one obtains that $\nu_{12}^{(0)}=(J_1,J_2)^{(2,0)}+(\b J_1,\b J_2)^{(0,2)}$. To prove that 
Leibnitz rule holds, one just needs to use the appropriate axiom of vertex operator algebra.\\  
The second point of the proposition immediately follows if one rewrites RHS of \rf{jac} via $N$-operation and uses the statements from 
point 1 of this proposition.$\blacksquare$ 

\vspace{3mm} 

\noindent
{\bf Remark.} Similar statement to Proposition 3.8. was given in \cite{malikov} in the context of the study of chiral de Rham complex. 

\section{Conclusions and Remarks}

In this paper we have studied the curved $\beta$-$\gamma$ system perturbed by the operator $g^{i\b j}p_ip_{\b j}$. One can also consider the general 
perturbation of conformal weight (1,1):
\begin{eqnarray}\label{gen}
V_{gen}=g^{i\b j}p_ip_{\b j}-\mu_{\bar{j}}^i p_i\p X^{\bar{j}}-\b \mu^{\bar{i}}_j p_{\bar i}\bp X^{j}+b_{i\b j}\p X^{i} \bp X^{\bar{j}}+hR^{(2)}(s)\t \Phi
\end{eqnarray} 
which, after integration over $p$-variables, transforms into the general string sigma model \rf{sigma}. The same way we associated with 
$g^{i\b j}p_ip_{\b j}$ the bivector field, one can associate with perturbation \rf{gen} an object from 
$\Gamma((T'M\oplus {T'}^*M)\otimes (T''M\oplus {T''}^*M)$. The conformal invariance condition at one loop will no longer be bilinear but 
we suggest that the equations will be described by means of the ``double commutator'' structure as in Definition 2.1. and its generalizations. 
Really, one can see that at the second order of perturbation theory in $V_{gen}$ the term $(V_{gen}, V_{gen})^{(1,1)}$ leads to double commutator, 
where the commutators are replaced by the Dorfman brackets.  

The formalism we developed in section 2, allows us in principle to give the cohomological meaning to the equations of conformal invariance, namely:
\begin{eqnarray}
[Q,\phi^{(0)}]+C_2(\phi^{(0)}, \phi^{(0)})+C_3(\phi^{(0)},\phi^{(0)},\phi^{(0)})+...=0,
\end{eqnarray} 
where $C_n$ are graded (with respect to the ghost number) multilinear operations, satisfying quadratic relations leading to $L_{\infty}$-like structure. 
In section 2 we described $C_2$ operation and its properties. We expect that the next nontrivial operation $C_3$ has the following form:
\begin{eqnarray}
C_3(\phi^{(0)},\phi^{(0)},\phi^{(0)})(z)=\mathcal{P}\int_{\mathcal{V}_{\epsilon,z}} \phi^{(1)}\phi^{(2)} \phi^{(0)}(z),
\end{eqnarray} 
where the three dimensional region $\mathcal{V}_{\epsilon,z}$ should depend on some parameter $\epsilon$ 
and $\mathcal{P}$ is the projection on the $\epsilon$-independent term as in Definition 3.1.

We mention also that in comparison to sigma model, studying the perturbed $\beta$-$\gamma$ system by means of conformal perturbation theory 
looks somewhat more promising since the underlying free CFT is the simplest possible. Again, in contrast to the usual sigma model throughout 
the perturbation theory one can keep geometric structures not destroyed.  

In the subsequent paper we will further develop the formalism related to $M$-operation and apply it directly to the 
``realistic'' sigma model \rf{sigma} describing strings in background fields. 

\subsection*{Acknowledgements}
The author is grateful to A.S. Losev for 
introduction in the subject and fruitful discussions. 
It is important to mention that the hypotheses concerning the using of generalized Maurer-Cartan equations and $L_{\infty}$-structures  
in the context of study of conditions of conformal invariance belong to A.S. Losev.
The author is very grateful to I.B. Frenkel, M. Kapranov and G. Zuckerman for numerous discussions on the subject 
and to I.B. Frenkel and N.Yu. Reshetikhin for their permanent encouragement and support.

\section*{Appendix A}
{\bf Proposition 2.1.} 
{\it The equations 
\begin{eqnarray}
&&\label{R}R^{\mu\nu}={1\over 4} H^{\mu\lambda\rho}H^{\nu}_{\lambda\rho}-2\nabla^{\mu}
\nabla^{\nu}\Phi,\\
&&\label{H}\nabla_{\mu}H^{\mu\nu\rho}-2(\nabla_{\lambda}\Phi)H^{\lambda\nu\rho}=0,
\end{eqnarray}
where metric, B-field, and a dilaton are expressed as follows: 
\begin{eqnarray}
G_{i\bar{k}}=g_{i\bar{k}}, \quad B_{i\bar{k}}=-g_{i\bar{k}}, \quad \Phi=\log\sqrt{g}+\Phi_0,
\end{eqnarray}
are equivalent to the following system:
\begin{eqnarray}
&&\p_i\p_{\bar{k}}\Phi_0=0,\quad \p_{\bar{p}}d^{\Phi_0}_{\bar{l}}g^{\bar{l}k}=0, 
\quad \p_{p}d^{\Phi_0}_{l}g^{\bar{k}l}=0,\nonumber\\
&&\label{gg}2g^{r\bar{l}}\p_r\p_{\bar{l}}g^{i\bar{k}}-2\p_r g^{i\bar{p}}\p_{\bar{p}}g^{r\bar{k}}-
g^{i\bar{l}}\p_{\bar{l}}d^{\Phi_0}_sg^{s\bar{k}}-g^{r\bar{k}}\p_r d^{\Phi_0}_{\bar{j}}g^{\bar{j}i}+\nonumber\\
&&\p_rg^{i\bar{k}}d^{\Phi_0}_{\bar{j}}g^{\bar{j}r}+\p_{\bar{p}}g^{\bar{k}i}d^{\Phi_0}_n g^{n\bar{p}}=0,
\end{eqnarray}
where $d^{\Phi_0}_ig^{i\bar{j}}\equiv \p_i g^{i\bar{j}}-2\p_i\Phi_0 g^{i\bar{j}}$ and $d^{\Phi_0}_{\b i}g^{\b i j}\equiv \p_{\b i} g^{j\bar{i}}-2\p_{\b i} \Phi_0 g^{j\bar{i}}$.\\
}
{\bf Proof.}We use the following formula for the Ricci tensor \cite{fock}:
\begin{eqnarray}
R^{\mu\nu}= 
1/2  G^{\alpha\beta}\partial_{\alpha}\partial_{\beta}G^{\mu\nu}+
\Gamma^{\mu\nu}-\Gamma^{\mu,\alpha\beta}\Gamma^{\nu}_{\alpha\beta},
\end{eqnarray}
where 
$$
\Gamma^{\mu\nu}=G^{\mu\rho}G^{\nu\sigma}\Gamma_{\rho\sigma},\quad
\Gamma_{\rho\sigma}=\half  (\partial_{\rho}\Gamma_{\sigma}+\partial_{\sigma}
\Gamma_{\rho})-\Gamma^{\nu}_{\rho\sigma}\Gamma_{\nu},
$$
$$
\Gamma_{\nu}=G^{\alpha\beta}\partial_{\beta}G_{\alpha\nu}-
\half \partial_{\nu}\log (G).
$$
Remembering that $G_{i\bar{k}}=g_{i\bar{k}}$ and $B_{i\bar{k}}=-g_{i\bar{k}}$, this leads to: 
\begin{eqnarray}\label{ricci}
&&R^{\mu\nu}-{1\over 4} H^{\mu\lambda\rho}H^{\nu}_{\lambda\rho}+2\nabla^{\mu}\nabla^{\nu}\Phi= \nonumber\\
&&-{1\over 4} H^{\mu\lambda\rho}H^{\nu}_{\lambda\rho}+
1/2  G^{\alpha\beta}\partial_{\alpha}\partial_{\beta}G^{\mu\nu}+
\t \Gamma^{\mu\nu}-\Gamma^{\mu,\alpha\beta}\Gamma^{\nu}_{\alpha\beta},
\end{eqnarray}
where 
$$
\t \Gamma_{\rho\sigma}=\half  (\partial_{\rho}\t \Gamma_{\sigma}+\partial_{\sigma}
\t \Gamma_{\rho})-\Gamma^{\nu}_{\rho\sigma}\t \Gamma_{\nu},
$$
$$ 
\t \Gamma_{\nu}=G^{\alpha\beta}\partial_{\beta}G_{\alpha\nu}+2\p_{\nu}\Phi_0.
$$
Here $\Phi_0=\Phi-\log \sqrt{g}$ and we denoted the determinant of matrix $g_{i\bar{j}}$ by $g$. 
Now let us study the third term in (\ref{ricci}): first,
for the components of $ \Gamma^{\nu}_{\alpha\beta}$, one has:
\begin{eqnarray}\label{crist}
&&\Gamma^{i}_{rs}=\half  g^{i\bar{k}}(\partial_{r}g_{\bar{k}s}+
\partial_{s}g_{\bar{k}r}),  \nonumber\\
&&\Gamma^{i}_{r\bar{s}}=\half g^{i\bar{k}}(\partial_{\bar{s}}g_{r\bar{k}}
-\partial_{\bar{k}}
g_{r\bar{s}}) \quad and \quad c.c.,
\end{eqnarray}
while all other components vanish. Therefore, one finds that
$\Gamma_{\bar{i},r\bar{s}}=\half H_{\bar{s}\bar{i}r}$, hence the third term in
(\ref{ricci}) provides the contribution of the $H^2$-type with an additional
term in $\Gamma\Gamma$ for $\mu=\bar{i}$ and $\nu=j$:
\begin{eqnarray}
&&\Gamma^{\bar{i},kl}\Gamma^{j}_{kl}=-{1\over 4} (g^{k\bar{r}}\partial_{\bar{r}}
g^{l\bar{i}}+g^{l\bar{r}}\partial_{\bar{r}}
g^{k\bar{i}})g^{j\bar{p}}(\partial_{k}g_{\bar{p}l}+\partial_{l}g_{\bar{p}k})=
\nonumber\\
&&-{1\over 4} (g^{k\bar{r}}\partial_{\bar{r}}
g^{l\bar{i}}-g^{l\bar{r}}\partial_{\bar{r}}
g^{k\bar{i}})g^{j\bar{p}}(\partial_{k}g_{\bar{p}l}-\partial_{l}g_{\bar{p}k})-
g^{k\bar{r}}\partial_{\bar{r}}
g^{l\bar{i}}g^{j\bar{p}}\partial_{l}g_{\bar{p}k}=\nonumber\\
&&-{1\over 4}H^{\bar{i}kl}H^{j}_{kl}+\partial_{\bar{r}}g^{\bar{i}k}
\partial_{k}g^{\bar{r}j}.
\end{eqnarray}
Thus we can see that the equations \rf{R} with inhomogeneous ($i$, $\b k$) components can be rewritten in such a way:
\begin{eqnarray}
&&R^{i\bar{k}}-{1\over 4} H^{i\lambda\rho}H^{\bar{k}}_{\lambda\rho}+
2\nabla^{i}\nabla^{\bar{k}}\Phi=\nonumber\\
&&g^{r\bar{l}}\p_r\p_{\bar{l}}g^{i\bar{k}}-\p_r g^{i\bar{p}}\p_{\bar{p}}g^{r\bar{k}}+
\frac{1}{2}(\nabla^i\t \Gamma^{\bar{k}}+
\nabla^{\bar{k}}\t \Gamma^i)=\nonumber\\
&&g^{r\bar{l}}\p_r\p_{\bar{l}}g^{i\bar{k}}-\p_r g^{i\bar{p}}\p_{\bar{p}}g^{r\bar{k}}-\frac{1}{2}
g^{i\bar{l}}\p_{\bar{l}}d^{\Phi_0}_sg^{s\bar{k}}-\frac{1}{2}g^{r\bar{k}}\p_r d^{\Phi_0}_{\bar{j}}g^{\bar{j}i}
+\nonumber\\
&&\frac{1}{2}\p_rg^{i\bar{k}}d^{\Phi_0}_{\bar{j}}g^{\bar{j}r}+\frac{1}{2}\p_{\bar{p}}g^{\bar{k}i}d^{\Phi_0}_n g^{n\bar{p}},
\end{eqnarray}
where we remind that $\t \Gamma^{\mu}=-d^{\Phi_0}_{\nu}G^{\nu\mu}=-\partial_{\nu}G^{\nu\mu}+2\p_{\nu}\Phi_0G^{\nu\mu}$, while 
for homogeneous components ($i$, $k$ and $\b i$, $\b k$) we get the following expression:
\begin{eqnarray}
&&R^{ij}-{1\over 4} H^{i\lambda\rho}H^{j}_{\lambda\rho}+
2\nabla^{i}\nabla^{j}\Phi=\nonumber\\
&&\frac{1}{2}(\nabla^i\t \Gamma^{j}+\nabla^{j}\t \Gamma^i)=-\frac{1}{2}(g^{i\bar{p}}\p_{\bar{p}}d^{\Phi_0}_{\bar{l}}g^{\bar{l}k}+
g^{k\bar{p}}\p_{\bar{p}}d^{\Phi_0}_{\bar{l}}g^{\bar{l}i}).
\end{eqnarray}
Hence the equations 
$R^{\mu\nu}-{1\over 4} H^{\mu\lambda\rho}H^{\nu}_{\lambda\rho}+2\nabla^{\mu}\nabla^{\nu}\Phi=0$ 
are equivalent to the equations on the bivector field $g^{i \b j}$:
\begin{eqnarray}\label{fR}
&&2g^{r\bar{l}}\p_r\p_{\bar{l}}g^{i\bar{k}}-2\p_r g^{i\bar{p}}\p_{\bar{p}}g^{r\bar{k}}-
g^{i\bar{l}}\p_{\bar{l}}d^{\Phi_0}_sg^{s\bar{k}}-g^{r\bar{k}}\p_r d^{\Phi_0}_{\bar{j}}g^{\bar{j}i}
+\nonumber\\
&&\p_rg^{i\bar{k}}d^{\Phi_0}_{\bar{j}}g^{\bar{j}r}+
\p_{\bar{p}}g^{\bar{k}i}d^{\Phi_0}_n g^{n\bar{p}}=0,\\
&&\label{sR}g^{i\bar{p}}\p_{\bar{p}}d^{\Phi_0}_{\bar{l}}g^{\bar{l}k}+
g^{k\bar{p}}\p_{\bar{p}}d^{\Phi_0}_{\bar{l}}g^{\bar{l}i}=0\quad and \quad c.c..
\end{eqnarray}
Now let's rewrite in a similar way the second series of equations, namely, the Maxwell-like equations for the B-field \rf{H}.
First of all we notice that by the antisymmetry of $H^{\mu\nu\rho}$ and the formula $\Gamma^{\lambda}_{\lambda\mu}=\frac{1}{2}\p_{\mu}\log(G)$
\begin{eqnarray}
\nabla_{\mu}H^{\mu\nu\rho}=\p_{\mu}H^{\mu\nu\rho}+\Gamma^{\lambda}_{\lambda\mu}H^{\xi\nu\rho}=\p_{\mu}(gH^{\mu\nu\rho}).
\end{eqnarray}
Hence the equation \rf{H} is equivalent to
\begin{eqnarray}\label{h}
\nabla_{\mu}(e^{-2\Phi_0}H^{\mu\nu\rho})=0.
\end{eqnarray}
Expressing $H$ in terms of $g^{i\b j}$ and multiplying \rf{h} on $e^{2\Phi_0}$, we arrive to the following system of equations: 
\begin{eqnarray}
&&\label{fH}2g^{r\bar{l}}\p_r\p_{\bar{l}}g^{i\bar{k}}-2\p_r g^{i\bar{p}}\p_{\bar{p}}g^{r\bar{k}}-
g^{i\bar{l}}e^{2\Phi_0}\p_s(e^{-2\Phi_0}\p_{\bar{l}}g^{s\bar{k}})\nonumber\\
&&-g^{r\bar{k}}e^{2\Phi_0}\p_{\bar{j}}(e^{-2\Phi_0}\p_r g^{\bar{j}i})
+\p_rg^{i\bar{k}}d^{\Phi_0}_{\bar{j}}g^{\bar{j}r}+
\p_{\bar{p}}g^{\bar{k}i}d^{\Phi_0}_n g^{n\bar{p}}=0,\\
&&\label{sH}e^{2\Phi_0}\p_{\bar{l}}(e^{-2\Phi_0}g^{i\bar{p}}\p_{\bar{p}}g^{\bar{l}k}-e^{-2\Phi_0}
g^{k\bar{p}}\p_{\bar{p}}g^{\bar{l}i})=0\quad and \quad c.c..
\end{eqnarray}
Comparing the equations \rf{fR} and \rf{fH} we get: 
\begin{eqnarray}
\p_i\p_{\bar k}\Phi_0=0, 
\end{eqnarray}
and then \rf{sR} and \rf{sH} lead to:
\begin{eqnarray}
\p_{\bar{p}}d^{\Phi_0}_{\bar{l}}g^{\bar{l}k}=0,\quad \p_{p}d^{\Phi_0}_{l}g^{l\b k}=0.
\end{eqnarray}
Thus the final system of the equations on $g^{i\b j}$ is:
\begin{eqnarray}
&&2g^{r\bar{l}}\p_r\p_{\bar{l}}g^{i\bar{k}}-2\p_r g^{i\bar{p}}\p_{\bar{p}}g^{r\bar{k}}-
g^{i\bar{l}}\p_{\bar{l}}d^{\Phi_0}_sg^{s\bar{k}}-g^{r\bar{k}}\p_r d^{\Phi_0}_{\bar{j}}g^{\bar{j}i}
+\nonumber\\
&&\p_rg^{i\bar{k}}d^{\Phi_0}_{\bar{j}}g^{\bar{j}r}+
\p_{\bar{p}}g^{\bar{k}i}d^{\Phi_0}_n g^{n\bar{p}}=0,\nonumber\\
&&\p_{\bar{p}}d^{\Phi_0}_{\bar{l}}g^{\bar{l}k}=0 \quad \p_{p}d^{\Phi_0}_{l}g^{l\b k}=0, \quad 
\p_i\p_{\bar{k}}\Phi_0=0.\quad \blacksquare
\end{eqnarray}

\vspace{3mm} 

\noindent
{\bf Proposition 2.2} {\it The equation
\begin{eqnarray}\label{Phi}
4(\nabla_{\mu}\Phi)^2-4\nabla_{\mu}\nabla^{\mu}\Phi+
R+{1\over 12} H_{\mu\nu\rho}H^{\mu\nu\rho}=0,
\end{eqnarray}
where metric, B-field, and dilaton are constrained by \rf{constr} and governed by  equations \rf{component}, is equivalent 
to the following one:
\begin{eqnarray}
d^{\Phi_0}_{i}d^{\Phi_0}_{\bar{j}}g^{\bar{j}i}=0,
\end{eqnarray}
where $d^{\Phi_0}_{i}d^{\Phi_0}_{\bar{j}}g^{\bar{j}i}\equiv (\p_{\b j}-2\p_{\b j}\Phi_0)(\p_i -2\p_i\Phi_0) g^{i\bar{j}}$.
}\\
{\bf Proof.} 
Using the equation \rf{R} one can reduce \rf{Phi} to 
\begin{eqnarray}\label{Phi2}
4(\nabla_{\mu}\Phi)^2-2\nabla_{\mu}\nabla^{\mu}\Phi-{1\over 6} H_{\mu\nu\rho}H^{\mu\nu\rho}=0.
\end{eqnarray}
Let's reexpress each term from the equation \rf{Phi2} by means of $g^{i\b j}$ and $\Phi_0$:
\begin{eqnarray}
&&4(\nabla_{\mu}\Phi)^2=2g^{i\b k}\p_i\log g\p_{\b k}\log g+ 4g^{i\b k}\p_i\log g\p_{\b k}\Phi_0\\
&&+4g^{i\b k}\p_{\b k}\log g\p_i\Phi_0+4g^{i\b k}\p_{\b k}\log g\p_i\Phi_0+8g^{i\b k}\p_i\Phi_0\p_{\b k}\Phi_0,\nonumber
\end{eqnarray}
\begin{eqnarray}
&&-2\nabla_{\mu}\nabla^{\mu}\Phi=-2G^{\mu\nu}\p_{\mu}\p_{\nu}\Phi+\Gamma^{\nu}\p_{\nu}\Phi=\\
&&-2G^{\mu\nu}\p_{\mu}\p_{\nu}\Phi-2G^{\mu\nu}\p_{\nu}\log g\p_{\mu}\Phi-2\p_{\mu}G^{\mu\nu}\p_{\nu}\Phi=\nonumber\\
&&-2g^{i\b k}\p_i\p_{\b k} \log g-\p_p g^{p\b r}\p_{\b r}\log g-\p_{\b s} g^{\b s k}\p_k \log g\nonumber\\
&&- 2 \p_p g^{p\b r}\p_{\b r}\Phi_0-2 \p_{\b s}g^{\b s k}\p_k\Phi_0-2g^{i\b k}\p_i\log g\p_{\b k}\log g\nonumber\\
&&-2g^{i\b k}\p_i\log g\p_{\b k}\Phi_0-2g^{i\b k}\p_{\b k}\log g\p_i\Phi_0,\nonumber
\end{eqnarray}
\begin{eqnarray}
&&-\frac{1}{6}H_{\mu\nu\rho}H^{\mu\nu\rho}=-H^{l m\b n}H_{l m\b n}=\\
&&(g^{m\b p}\p_{\b p} g^{l\b n}- g^{l\b r}\p_{\b r} g^{m\b n})(\p_l g_{m\b n}-\p_m g_{l\b n})=\nonumber\\
&&2g^{l\b r}\p_{\b r}g^{m\b n}\p_m g_{l\b n}-2g^{l\b r}\p_{\b r}g^{m\b n}\p_lg_{m\b n}.\nonumber
\end{eqnarray}
Now rewriting \rf{Phi} we get:
\begin{eqnarray}
&&-2g^{i\b k}\p_i\p_{\b k} \log g-\p_p g^{p\b r}\p_{\b r}\log g-\p_{\b s} g^{\b s k}\p_k \log g\\
&&-2g^{i\b k}\p_i\log g\p_{\b k}\Phi_0-2g^{i\b k}\p_{\b k}\log g\p_i\Phi_0+2g^{i\b k}\p_{\b k}\log g\p_i\Phi_0\nonumber\\
&&+2g^{i\b k}\p_{\b k}\log g\p_i\Phi_0+8g^{i\b k}\p_i\Phi_0\p_{\b k}\Phi_0=0.\nonumber
\end{eqnarray}
Let's consider the sum of this equation with \rf{gg} contracted with $-g_{i\b k}$:
\begin{eqnarray}   
&&-2g^{r\bar{l}}\p_r\p_{\bar{l}}g^{i\bar{k}}g_{i\bar{k}}+2\p_r g^{i\bar{p}}\p_{\bar{p}}g^{r\bar{k}}g_{i\bar{k}}+
\p_{\bar{k}}d^{\Phi_0}_sg^{s\bar{k}}+\p_i d^{\Phi_0}_{\bar{j}}g^{\bar{j}i}+\nonumber\\
&&\p_r\log gd^{\Phi_0}_{\bar{j}}g^{\bar{j}r}
+\p_{\bar{p}}\log g d^{\Phi_0}_n g^{n\bar{p}}=0.
\end{eqnarray}
In such a way, using the formula $\p_{\mu}g=g^{i\b k}\p_{\mu}g_{i\b k}$ we get:
\begin{eqnarray}  
8g^{i\b k}\p_i\Phi_0\p_{\b k}\Phi_0+
\p_{\bar{k}}d^{\Phi_0}_sg^{s\bar{k}}+\p_i d^{\Phi_0}_{\bar{j}}g^{\bar{j}i}=0,
\end{eqnarray}
which can be rewritten as:
\begin{eqnarray} 
d^{\Phi_0}_{i}d^{\Phi_0}_{\bar{j}}g^{\bar{j}i}=0.\quad \blacksquare 
\end{eqnarray}

\section*{Appendix B}
{\bf Proposition 3.3.} {\it Let $\phi^{(0)}$ be the element of $S^0$, satisfying the following conditions:
$\phi^{(2)}=dz\wedge d\b z h^{-1}g^{i\b j}(X, \b X)p_ip_{\b j}$, the dilatonic terms $U(X,\b X)=b_{1}b_{-1}\phi^{(0)}$, and 
$\b U(X, \b X)=\t b_{1}\t b_{-1}\phi^{(0)}$ are respectively functions of $X^i(z)$,  $X^{\b i}(\b z)$. Then \rf{mc}  
gives the equations \rf{comp} and \rf{dil} expanded up to the second order in the formal parameter t, 
such that the expansion of matter fields is: $g^{i\b j}=tg_1^{i\b j}+t^2g_2^{i\b j}+...$, 
$\Phi_0=1/2(-\log(\omega\b \omega)+(tU_1+t\b U_1)+t^2(U_2+\b U_2)+...)$.}\\
{\bf Proof.} 
The first equation of \rf{mc} leads to the  
following relations between $U$-, $W$-, and $V$- terms:
\begin{eqnarray}\label{m10}
&&(L_0V_1)(z)-V_1(z)+L_{-1}\b W_1+ \b L_{-1} W_1=0, \nonumber\\
&&W_1=-1/2((\b L_1 V_1)+L_{-1}\b U_1), \quad \b W_1=-1/2((L_1 V_1)+\b L_{-1}U_1),\nonumber\\
&&L_1W_1=0, \b L_1\b W_1=0.
\end{eqnarray}
Therefore it is clear that 
\begin{eqnarray}
W_1=1/2d_{\b j}^{\t \Phi}g_1^{i\b j}p_i-1/2\p X^i\p_i\b U , \quad 
 \b W_1=1/2d_i^{\t \Phi}g_1^{i\b j}p_{\b j}-1/2\bp X^{\b i}\p_{\b i}U, 
\end{eqnarray}
where $\t \Phi=-1/2\log(\omega+\b \omega)$, and hence \rf{m10} gives the equations:
\begin{eqnarray}\label{lin}
\p_i\p_{\b k}(U_1+\b U_1)=0, \quad \p_{\b k}d_{\b j}^{\t \Phi}g_1^{i\b j}=0, \quad
\p_k d_i^{\t \Phi}g_1^{i\b j}=0, \quad
d_{\b j}^{\t \Phi}d_i^{\t \Phi}g_1^{i\b j}=0,
\end{eqnarray}
which coincide with \rf{comp}, \rf{dil} at the first order in $t$.\\ 
The second equation of \rf{mc} leads to more complicated conditions:
\begin{eqnarray}\label{m20}
&&(L_0V_2)-V_2-1/2(V_1,V_1)^{(1,1)}+1/2
( \b W_1,V_1)^{(0,1)}-\nonumber\\
&&1/2
( V_1,\b W_1)^{(0,1)}
+1/2(W_1,V_1)^{(1,0)}-\nonumber\\
&&1/2(V_1,W_1)^{(1,0)}+L_{-1}\b W_2+ \b L_{-1} W_2=0,
\end{eqnarray}
\begin{eqnarray}
&&\label{W}\b W_2=-1/2((L_1 V_2)-(V_1,V_1)^{(2,1)}+(\b W_1,V_1)^{(1,1)}\nonumber\\
&&+(W_1,V_1)^{(2,0)}+1/2(U_1,V_1)^{(1,0)}-
1/2(V_1,U_1)^{(1,0)}+\b L_{-1}U_2),\nonumber\\
&&W_2=-1/2((\b L_1 V_2)-(V_1,V_1)^{(1,2)}+(W_1,V_1)^{(1,1)}+\nonumber\\
&&(\b W_1,V_1)^{(0,2)}+1/2(\b U_1,V_1)^{(0,1)}-
1/2(V_1,\b U_1)^{(0,1)}+L_{-1}\b U_2),
\end{eqnarray}
\begin{eqnarray}
&&\label{dilw}2L_1W_2-2L_0U_2+(U_1,W_1)^{(1,0)}-(W_1,U_1)^{(1,0)}+2(W_1,W_1)^{(2,0)}\\
&&-(V_1,W_1)^{(2,1)}+(V_1,U_1)^{(1,1)}+2(\b W_1,W_1)^{(1,1)}-(\b W_1,U_1)^{(0,1)}=0\nonumber\\
&&2(\b L_1\b W_2)-2\b L_0\b U_2+(\b U_1,\b W_1)^{(0,1)}-(\b W_1,\b U_1)^{(0,1)}+2(\b W_1,\b W_1)^{(0,2)}\nonumber\\
&&-(V_1,\b W_1)^{(1,2)}+(V_1,\b U_1)^{(1,1)}+2(W_1,\b W_1)^{(1,1)}-(W_1,\b U_1)^{(0,1)}=0\nonumber.
\end{eqnarray}
The equations \rf{W} give the following expression for $W_2$ and $\b W_2$:
\begin{eqnarray}\label{W2}
\b W_2=-1/2(-d_i^{\t \Phi}g_2^{i\b j}p_{\b j}+\p_i(U_1+1/2\b U_1)g_1^{i\b j}p_{\b j}+\bp X^{\b i} \p_{\b i}U_2)+O(h),\nonumber\\
W_2=-1/2(-d_{\b j}^{\t \Phi}g_2^{i\b j}p_{i}+\p_{\b i}(\b U_1+1/2 U_1)g_1^{\b i j}p_{j}+\p X^i\p_i\b U_1)+O(h),
\end{eqnarray}
Let's substitute the expressions for $W_2, \b W_2$ into \rf{m20}. 
The coefficient of $\p X^i \bp X^{\b j}$  gives the familiar equation 
\begin{eqnarray}\label{comp1}
\p_i\p_{\b k}(U_2+\b U_2)=0,
\end{eqnarray}
  while coefficients of $p_i\bp X^{\b j}$, $p_{\b i}\p X^{j}$,  and
$p_ip_{\b j}$ in \rf{m20} at the zeroth order in $h$ lead to the following:
\begin{eqnarray}\label{comp2}
&&\p_{\bar{p}}(d^{\t \Phi}_{\bar{l}}g_2^{\bar{l}k}-\p_{\bar{l}}(U_1+\b U_1)g_1^{\bar{l}k})=0, 
\quad \p_{p}(d^{\Phi_0}_{l}g_2^{\bar{k}l}-\p_{l}(U_1+\b U_1)g_1^{\bar{k}l})=0,\nonumber\\
&&2g_1^{r\bar{l}}\p_r\p_{\bar{l}}g_1^{i\bar{k}}-2\p_r g_1^{i\bar{p}}\p_{\bar{p}}g_1^{r\bar{k}}-
g_1^{i\bar{l}}\p_{\bar{l}}d^{\t \Phi}_sg_1^{s\bar{k}}-g_1^{r\bar{k}}\p_r d^{\t \Phi}_{\bar{j}}g_1^{\bar{j}i}+\nonumber\\
&&\p_rg_1^{i\bar{k}}d^{\t \Phi}_{\bar{j}}g_1^{\bar{j}r}+\p_{\bar{p}}g_1^{\bar{k}i}d^{\t \Phi}_n g_1^{n\bar{p}}=0.
\end{eqnarray}
The equations \rf{comp1}, \rf{comp2} coincide with \rf{comp} at the order $t^2$.
Let's look on the equations \rf{dilw} which we did not touch before. We see that at the first order in $h$ they both lead to the following relation:     
\begin{eqnarray}
&&-d^{\t \Phi}_id_{\b j}^{\t \Phi}g_2^{i\b j}+3/2g_1^{i\b j}\p_i\p_{\b j}(U_1+\b U_1)\nonumber\\
&&+\p_i(U_1+\b U_1)d_{\b j}^{\t \Phi}g_1^{i\b j}+
\p_{\b j}(U_1+\b U_1)d_{i}^{\t \Phi}g_1^{i\b j}=0,
\end{eqnarray}
which by means of condition \rf{lin} is equivalent to:
\begin{eqnarray}
-d^{\t \Phi}_id_{\b j}^{\t \Phi}g_2^{i\b j}+\p_i(U_1+\b U_1)d_{\b j}^{\t \Phi}g_1^{i\b j}+
\p_{\b j}(U_1+\b U_1)d_{i}^{\t \Phi}g_1^{i\b j}=0.
\end{eqnarray}
Using the relation between $\t \Phi$, $\Phi_0$, and $\Phi$ we see that this equation coincides with \rf{dil} at the order $t^2$. $\blacksquare$

\end{document}